\documentclass[12pt]{article}
\usepackage{amsfonts}
\usepackage{amssymb}
\usepackage{cite}

\def\ep{\epsilon}

\def\Tr{{\rm Tr} \,\! }

\def\cdots {\cdot\cdot\cdot}
\def\pint#1 {- \!\!\!\!\!\!\!\! \,\int_{#1}}

\def\ni       {\noindent}
\def\lb       {\left( }
\def\rb       {\right) }

\def\comma      { \, , }
\def\period     { \, . }
\def\semiket#1  { \, #1 \, \rangle \, }
\def\del        {  \partial  }

\def\half       {  {1\over 2}  }
\def\abs#1      {  \, \vert #1 \vert \,   }
\def\Im#1    { \, {\rm Im } \, #1  }
\def\Re#1    { \, {\rm Re}  \, #1  }

\def\binom#1#2 { \vecii{ {}_{#1} }{\raisebox{.5ex}{$ {}^{#2} $}} }
\def\sqbinom#1#2 { \Bigl(\begin{array}{c} {}_{#1}
                       \\ \raisebox{.5ex}{${}^{#2}$} \end{array}\Bigr)^2  }

\def\bfZ     { {\bf Z}}

\def\zbar   {\bar{z}}
\def\wbar   {\bar{w}}

\def\r12    {\frac{r_1}{r_2}}
\def\calD  {{\cal D}}
\def\calO  {{\cal O}}
\def\calR {{\cal R}}

\def\tilalpha   {\tilde{\alpha}}

\def\tilL   {\tilde{L}}

\def\CFT { {\rm CFT} }

\def\tC      {{}^{t}\!C}
\def\vecii#1#2      {  \Biggl(\! \begin{array}{c}#1\\#2\end{array}\!\Biggr)  }
\def\veciii#1#2#3   {  \left(\begin{array}{c}#1\\#2\\#3\end{array}\right)  }
\def\matrixii#1#2#3#4            {  \Biggl( \! \begin{array}{cc}#1&#2\\#3&#4
                                       \end{array} \! \Biggr) }
\def\matrixiii#1#2#3#4#5#6#7#8#9 {  \left(\begin{array}{ccc}#1&#2&#3\\
                                     #4&#5&#6\\#7&#8&#9\end{array}\right)  }
\def\eqb         {  \begin{eqnarray}  }
\def\eqe           {  \end{eqnarray}  }
\def\nn               {  \nonumber  }

\def\appendixnumbering { \setcounter{equation}{0}
         \renewcommand{\theequation}{\Alph{section}.\arabic{equation}}}
\def\msection#1{ \addtocounter{section}{1} \setcounter{subsection}{0}
   \par \bigskip
      \par \bigskip \noindent
   {\large\bf \arabic{section} \quad  #1 }
    \par \bigskip}
\def\appsection#1{\setcounter{section}{1} \setcounter{subsection}{0}
                 \appendixnumbering
      \par \bigskip \par\bigskip\noindent
   {\large\bf  #1 }
    \par \bigskip}
\def\msubsection#1{\addtocounter{subsection}{1}
      \noindent  {\normalsize\it
      \arabic{section}.\arabic{subsection} \quad #1  }
   \par \medskip }

\def\csectionast#1    { \begin{center}
    {\large\bf #1  }   \end{center} \par \bigskip}
\def\titleandfile#1#2   {  \begin{center}{\large\bf #1}\end{center}
                            \par\begin{flushright} #2 \end{flushright} 
                            \par \begin{flushright} \today \end{flushright}}

\renewcommand{\thefootnote}{\fnsymbol{footnote}}

\begin{document}
\def\papertitlepage{\baselineskip 3.5ex \thispagestyle{empty}}
\def\preprinumber#1#2#3{\hfill \begin{minipage}{2.6cm} #1
                \par\noindent #2
              \par\noindent #3
             \end{minipage}}
\renewcommand{\thefootnote}{\fnsymbol{footnote}}
%
%
\papertitlepage
\setcounter{page}{0}
\preprinumber{}{UTHEP-570}{arXiv:0809.4548}
\baselineskip 0.8cm
\vspace*{2.0cm}
\begin{center}
{\large\bf Entanglement through conformal interfaces}
\end{center}
\vskip 4ex
\baselineskip 1.0cm
\begin{center}
        { Kazuhiro ~Sakai\footnote[2]{\tt sakai@phys-h.keio.ac.jp}, } \\
 \vskip -1ex
    {\it Department of Physics, Keio University}
 \vskip -2ex   
    {\it Hiyoshi, Yokohama 223-8521, Japan} \\
 \vskip 2ex
     { Yuji  ~Satoh\footnote[3]{\tt ysatoh@het.ph.tsukuba.ac.jp}}  \\
 \vskip -1ex
    {\it Institute of Physics, University of Tsukuba} \\
 \vskip -2ex
   {\it Tsukuba, Ibaraki 305-8571, Japan}
\end{center}
\vskip 12ex
%
\baselineskip=3.5ex
\begin{center} {\bf Abstract} \end{center}

\par\medskip
\ 
We consider entanglement through permeable  interfaces
in the $c=1$ (1+1)-dimensional conformal field theory.  
We compute the partition functions with the interfaces inserted. 
By the replica trick, the entanglement entropy is obtained analytically.
The entropy scales logarithmically with respect to the size of the system,  
similarly to the universal scaling of the ordinary entanglement entropy 
in (1+1)-dimensional conformal field theory.
Its coefficient, however, is not constant but controlled by the
permeability, the dependence on which is expressed through the dilogarithm 
function. The sub-leading term of the entropy counts the winding numbers, 
showing an analogy to the topological entanglement entropy which characterizes 
the topological order in (2+1)-dimensional systems.
%
%
%
%
%

\vspace*{\fill}
\ni
September 2008

\newpage
\renewcommand{\thefootnote}{\arabic{footnote}}
\setcounter{footnote}{0}
\setcounter{section}{0}
\baselineskip = 3.3ex
\pagestyle{plain}
\msection{Introduction}

Conformal interfaces provide a natural framework to extend the 
(1+1)-dimensional conformal field theory (CFT) with boundaries. 
Taking into account the important roles played
by  the boundary  CFT  in condensed matter physics and string theory, 
one expects that the interface CFT also opens up new directions in such studies. 
In fact,  interesting properties have already been found: For example, all the symmetries 
of the rational CFT are generated by a class of interfaces called topological interfaces 
\cite{Frohlich:2004ef}. Topological interfaces also transform one set of D-branes to 
another \cite{Graham:2003nc,Bachas:2004sy}.
For (potential) applications and a (partial) list of references, we refer to 
\cite{Frohlich:2004ef,Graham:2003nc,Bachas:2004sy,Oshikawa:1996dj,Petkova:2000ip,
Bachas:2001vj,Quella:2002ct,Recknagel:2002qq,Quella:2006de,Fuchs:2007tx,Bachas:2007td,
Brunner:2008fa,Gang:2008sz} and references therein.
  
When a system consists of two (or more) sub-systems as in the interface CFT, 
the quantum correlation, i.e., entanglement,  between the sub-systems 
is a useful probe to the system. The entanglement entropy is a measure of this 
entanglement. In (1+1)-dimensional systems, the entanglement entropy of the ground 
state shows a sharp contrast between the critical and the non-critical regime 
\cite{Vidal:2002rm} and, at the critical point, there appears 
a universal logarithmic scaling with respect to the size of the system
characterized by the central charge \cite{Holzhey:1994we,Calabrese:2004eu}. 
In (2+1)-dimensional systems, the leading term of the ground state entanglement entropy 
scales linearly as the boundary size of the system, whereas the sub-leading term 
characterizes the topological order of the systems, named as the topological entanglement 
entropy \cite{Kitaev:2005dm,LW}. Entanglement is essential also in quantum computation 
and information. 
The references on the entanglement entropy in (1+1)-dimensional 
systems with defects include \cite{Levine, Peschel,SCLA}.
 
In this note, we consider the entanglement entropy in the $c=1$ interface CFT. 
A class of conformal interfaces in this theory has been constructed which interpolates
perfectly transmitting and reflecting interfaces \cite{Bachas:2001vj}.
In addition to  the physical relevance of the $c=1$ CFT, because of the fact that
these permeable interfaces are simple but possessing structures characterized by
some parameters, we expect that they provide useful insights into the interface CFT. 
The entanglement entropy may be a useful probe, and its role in the interface
CFT is of  interest. 
  
In section 2, we briefly summarize the entanglement entropy with slight
generalizations in the case of the interface CFT. We also introduce the $c=1$ 
permeable interfaces. In section 3, we compute the partition functions with the 
interfaces inserted. In section 4, using this result and the replica trick, we  
obtain the entanglement entropy analytically when the sizes of the two CFT's 
joining at the interface are equal. 
The entropy scales logarithmically with respect to 
the size of the system, similarly to the universal scaling in the case without interfaces. 
However, its coefficient is not constant but controlled by the permeability. 
The sub-leading term counts the product of the winding numbers. 
This shows an analogy to the topological entanglement entropy in (2+1)-dimensional 
systems. 
In deriving the entropy, we adopt two approaches: One is based on 
the Bernoulli polynomials and
numbers, which provides a general method for carrying out the replica trick.
Another is a direct evaluation of a sum by an integral for the large size of the system.
In the course of verifying the equivalence of the results from the two approaches,
we find that the scaling coefficient  is expressed by the dilogarithm function.   
In section 5, we conclude with a 
brief summary. Some useful formulas are collected in Appendix. 
The entanglement entropy in the $c=1$ interface CFT has been 
discussed in \cite{Azeyanagi:2007qj} in a different setting from ours and 
in the context of the boundary entropy and its holographic dual. 
\msection{Setup}
\msubsection{entanglement entropy in interface CFT}
We consider two (1+1)-dimensional CFT's defined on a half complex plane 
Re$ \, w > 0 $ and Re$\, w < 0$, respectively. The interface between $\CFT_{1}$ and 
$\CFT_{2}$ lies along the imaginary axis Re$\, w = 0$. The conformal invariance requires
the continuity condition
\eqb
  L_{n}^{1} - \tilL_{-n}^{1} =  L_{n}^{2} - \tilL_{-n}^{2} \comma \label{intconf}
\eqe
at the interface, where $L^{a}_{n}, \tilL_{n}^{a}$ $(a=1,2; \, n \in \bfZ)$ are
the left and the right Virasoro generators of $\CFT_{a}$.

We are interested in the  entanglement entropy of the ground state.
The entropy is defined by the von Neumann entropy of the reduced density matrix
for the ground state
$\rho_{1} = \Tr_{2} \, |0 \rangle \langle 0 |$ as
\eqb
   S = - \Tr_{1} \, \rho_{1} \log \rho_{1} = - \frac{\del}{\del K} \Tr_{1} \, \rho^{K}_{1} 
   \ \Big\vert_{K=1}
   \comma
\eqe
where $\Tr_{a}$ stands for the trace over the degrees of freedom in $\CFT_{a}$.
The trace of the $K$-th power of the reduced density matrix is represented by 
a partition function on a $K$-sheeted Riemann surface
$\calR_{K}$ whose branch cut runs  along the real axis
from $w=0$ to $\infty$
\cite{Callan:1994py}: 
\eqb
    \Tr_{1} \, \rho^{K}_{1} = \frac{Z(K)}{Z^{K}(1)} \equiv
    \frac{1}{Z^{K}(1)} \int \calD \phi \, \exp\Bigl[ - \int_{\calR_{K}}
    d^{2}w \, {\cal L}(\phi)
     \Bigr] \comma \label{Trrhon}
\eqe
where $\phi$ represents the fields of $\CFT_{1}$ and $\CFT_{2}$.
The interface is inserted on each sheet of $\calR_{K}$.
The normalization factor $1/Z^{K}(1)$ assures $\Tr_{1} \, \rho_{1} = 1$.
{}From this path-integral representation, the entropy is given by 
\eqb
   S = (1- \del_{K}) \, \log Z(K) \, \Big\vert_{K=1} \period \label{SZK}
\eqe
As in the case without interfaces, how mixed the reduced density matrix is depends 
on the correlation across the  interface, and hence the von Neumann entropy
measures the entanglement of the CFT's through the interface. 

To evaluate the partition function with the interface inserted, we move to 
$z =  \log w$ plane. 
Introducing cutoffs at $|w| = \epsilon$ and $|w| = L$ \cite{Holzhey:1994we,Levine},
the $K$-sheeted Riemann surface $\calR_{K}$ is mapped to a rectangular 
whose lengths along the real and the imaginary axis are
$(\log (L/\epsilon), 2\pi  K)$, repsectively.
The interface is mapped to Im$\, z = (2m-1) \pi /2$ $(m=1, ..., 2K)$. 
In the following, we set $\epsilon = 1/L$ for simplicity. 

Here, we impose the periodic boundary condition along the Re$\, z$ direction.
When the interface is absent, this reduces to the treatment in \cite{Holzhey:1994we}, 
and we can check that our final results actually reproduce those in \cite{Holzhey:1994we}
as a special case.  (Our $L$ corresponds to $\Sigma/\epsilon$ in \cite{Holzhey:1994we}, 
as is clear from our setting.)
Also, when the interface is topological (see below), our partition
function is regarded as a generalization 
(in the $c=1$ case)
of the generalized twisted partition functions discussed in \cite{Petkova:2000ip}.
In order to carry out the analysis independently of the boundary conditions,
it is desirable to generalize the analysis in 
\cite{Holzhey:1994we,Calabrese:2004eu,Cardy:1988tk} 
based on the conformal invariance to our case with interfaces.

Now, our partition function is a torus partition function with $2K$ interfaces 
inserted. Recall that the ground state density matrix is represented by 
the product of the ground state wave function $\Psi_{0} \Psi^{\ast}_{0}$
in deriving the path-integral representation (\ref{Trrhon}), and that
$\Psi_{0}$ ($\Psi^{\ast}_{0}$) gives the path-integral on the lower (upper) half 
$w$-plane. One then finds that
the interfaces at Im$ \, z = (2m-1)\pi  /2$ for odd $m$
and those for even $m$ are hermitian conjugate with each other. Therefore, 
we find that
\eqb
   Z(K) = \Tr_{1} \Bigl( I_{12} \, q^{L^{2}_{0}+\tilL^{2}_{0}} \, (I_{12})^{\dagger}\, 
   q^{L^{1}_{0}+\tilL^{1}_{0}} \, I_{12}
   \cdots (I_{12})^{\dagger} \, q^{L^{1}_{0}+\tilL^{1}_{0}} \Bigr) \comma \label{Zn}
\eqe
where $q = e^{-2\pi t}$ with $ t = \pi /(2 \log L) $, 
we have denoted the interfaces by $I_{12}$ and $(I_{12})^{\dagger}$, and
each of them appears $K$ times alternately. We have also rescaled 
the rectangular in the $z$-plane so that it becomes of the standard 
lengths $(2\pi, 2\pi t)$.
The subscript
``12'' is put to make explicit the fact that $I_{12}$ joins $\CFT_{1}$ and 
$\CFT_{2}$ in this order. The above expression is manifestly symmetric 
with respect to $\CFT_{1}$ and $\CFT_{2}$, as it should be.

\par\medskip
\par\medskip
\msubsection{$c=1$ permeable interfaces}
We now specialize our discussion to the case of the $c=1$ permeable
interfaces.  A general way to construct conformal interfaces is to use
the folding trick \cite{Oshikawa:1996dj}: Let us set
$\tilde{\tau} =$ Re$\, w$. 
Then, by flipping the sign of $\tilde{\tau}$ for $\tilde{\tau} < 0$, 
$\CFT_{2}$ comes to live on the $\tilde{\tau} > 0$ half plane. In the course,  
the interface becomes a conformal boundary, which can be expressed by a 
conformal boundary state of the tensor product theory $\CFT_{1} \otimes \CFT_{2}$.
In fact, the condition of the conformal invariance of the interface (\ref{intconf})
becomes that of the conformal boundary, 
$L_{n}^{1} + L_{n}^{2}- (\tilL_{-n}^{1}+ \tilL_{-n}^{2}) =  0$,
since the left and the right movers and the positive and the negative modes 
in $\CFT_{2}$ are exchanged, respectively, by the folding.
Conversely, unfolding a conformal boundary state gives a conformal
interface.

In this way, from the boundary states of the tensor product theory of two free bosons,
the $c=1$ permeable interfaces are obtained \cite{Bachas:2001vj}:
\eqb
   &&  I^{\pm \ (\alpha,\beta)}_{12\, (k_{1}, k_{2})} (\theta_{\pm})  =
   G^{\pm \ (\alpha,\beta)}_{12\, (k_{1}, k_{2})} (\theta_{\pm})
    \prod_{n=1}^{\infty} e^{
   \frac{1}{n} \Bigl( S^{\pm}_{11}\, \alpha_{-n}^{1} \tilalpha_{-n}^{1}
   -S^{\pm}_{12} \, \alpha_{-n}^{1} \alpha_{n}^{2}
   -S^{\pm}_{21}\, \tilalpha_{-n}^{1} \tilalpha_{n}^{2}
   +S^{\pm}_{22}\, \alpha_{n}^{2}\tilalpha_{n}^{2}
   \Bigr)
   }  \comma \nn \\
    && \qquad \qquad G^{+ \ (\alpha,\beta)}_{12\, (k_{1}, k_{2})} (\theta_{+})
    =  g_{+} \sum_{N,M=-\infty}^{\infty} e^{ i(N\alpha - M \beta)}
   |k_{2} N,k_{1}M\rangle \langle k_{1}N,k_{2}M | \comma \label{I} \\
    && \qquad \qquad G^{- \ (\alpha,\beta)}_{12\, (k_{1}, k_{2})} (\theta_{-})
        =  g_{-} \sum_{N,M=-\infty}^{\infty} e^{ i(N\alpha - M \beta)}
   |k_{1} M,k_{2}N\rangle \langle k_{1}N,k_{2}M | \comma \nn
\eqe
where
\eqb
   S^{\pm} 
  = \matrixii{\mp\cos 2\theta_{\pm}}{-\sin 2\theta_{\pm}}{\mp\sin 2 \theta_{\pm}}{\cos 2 \theta_{\pm}} \comma \qquad
    g_{\pm} 
      = \Big\vert \frac{k_{1}k_{2}}{\sin 2 \theta_{\pm}} \Big\vert^{1/2} \comma
      \label{Sg}
\eqe
and
\eqb
   \tan \theta_{+} = \frac{k_{2}R_{2}}{k_{1}R_{1}} \comma \quad
   \tan \theta_{-} = \frac{2k_{2} R_{1}R_{2}}{k_{1}} \period \label{kRtheta}
\eqe
$\alpha_{n}^{a}, \tilalpha^{a}_{n}$ $(a= 1,2 ; \, n \in \bfZ)$  are the modes of
the free boson $\phi_{a}$ compactified on a circle with radius $R_{a}$.
They satisfy $[\alpha^{a}_{m}, \alpha^{b}_{n}] =  m \delta_{m,n}\delta^{ab}$ and
similar expressions for $\tilalpha^{a}_{n}$. It is understood that
$\alpha^{a}_{-n}, \tilalpha^{a}_{-n}$ $(n>0)$ implicitly act on $G^{\pm}_{12}$ from the left
and $\alpha^{a}_{n}, \tilalpha^{a}_{n}$ $(n>0)$ from the right.
$k_{a}$ is the winding number for $\phi_{a}$. 
$\vert n_{a}, m_{a} \rangle$
is the oscillator vacuum for $\phi_{a}$ with the momentum $n_{a}/R_{a}$ and 
the winding number $m_{a}$. Its dual is denoted by $\langle n_{a}, m_{a} \vert$.
$I_{12}^{+}$ is obtained from 
the boundary state with one Dirichlet and one Neumann boundary condition. $\theta_{+}$
is the angle between the $\phi_{1}$ and the Neumann direction in the target space.
$I_{12}^{-}$ is obtained by T-dualizing the boundary state for $I_{12}^{+}$.  
In unfolding, there is a choice of which CFT is unfolded. Changing this choice
gives ``anti-interfaces'' $\bar{I}_{21}^{\pm}$ \cite{Bachas:2007td},
which are equivalent to $(I_{12}^{+})^{\dagger}$ with some signs of
the parameters flipped.
One can  check explicitly that the continuity condition (\ref{intconf}) is
satisfied by these interfaces.

The matrices $S^{\pm}$ control  interactions between $\CFT_{1}$ and $\CFT_{2}$.
When $\theta_{\pm} $ are a multiple of $\pi/2$, the two CFT's decouple and the 
interfaces become totally reflective. 
When $\theta_{\pm}$ are odd multiple of 
$\pi/4$, the interfaces become totally transmissive.  In this case, 
the interface is called topological, since
each of the left and the right energy-monentum tensor becomes continuous
across the intreface, i.e., $L^{1}_{n} = L^{2}_{n}$ and $\tilL^{1}_{n} = \tilL^{2}_{n}$,
and hence the interface can be freely deformed. 
Note that the identity operator is included as a special
case of the topological interface $I^{+ \ (0,0)}_{12\, (1,1)} (\pi/4)$.
%
\msection{Partition functions with interfaces inserted}
In this section, we compute the partition function with the interfaces
inserted, i.e., $Z(K)$
in (\ref{Zn}) for $I^{\pm}_{12}$. The computation  in the following can be regarded
as a generalization of (part of) that in \cite{Bachas:2007td} for the fusion of the interfaces. 

\par\medskip\par\medskip
\msubsection{case of $I^{+}_{12}$}

Here, we consider the case of $I^{+}_{12}$.
To carry out the computation, it is convenient to first focus on a unit of the products
of the operators in $Z(K)$,
\eqb
   J  =  I^{+}_{12} \, q^{L_{0}^{2} + \tilL_{0}^{2}} \, (I^{+}_{12})^{\dagger}\,  
   q^{L_{0}^{1} + \tilL_{0}^{1}} \label{J}
   \comma
\eqe
and rewrite the quadratic oscillator parts in $I^{+}_{12}$ and $(I^{+}_{12})^{\dagger}$ as
\eqb
   ({\alpha_{-n}^{1}}, {\tilalpha_{n}^{2}}) \cdot \matrixii{-c}{s}{s}{c} 
     \cdot \vecii{\tilalpha_{-n}^{1}}{\alpha_{n}^{2}}  
   \comma \quad 
   ({\alpha_{-n}^{2}}, {\tilalpha_{n}^{1}}) \cdot \matrixii{c}{s}{s}{-c}
    \cdot \vecii{\tilalpha_{-n}^{2}}{\alpha_{n}^{1}} 
   \comma
\eqe
respectively, where $c = \cos 2 \theta_{+} \comma  s = \sin 2 \theta_{+}$.
We then linearize the quadratic forms by an identity
\eqb
  e^{\vec{A} \cdot \vec{B}} = \int \frac{d^{2} \vec{z}}{\pi^{2}} \, e^{-\vec{z}\cdot \vec{\zbar} 
   - \vec{z}\cdot \vec{A} - \vec{\zbar}\cdot \vec{B}} \comma
\eqe
which is valid when all $A_{i}, B_{i}$ $(i=1,2)$ are commuting with each other. 
After the linearization,  one can explicitly put the creation operators 
$\alpha^{a}_{-n}, \tilalpha^{a}_{-n}$ $(n>0)$ on the left of 
$G^{+}_{12}$ or $(G^{+}_{12})^{\dagger}$,
and the annihilation operators $\alpha^{a}_{n}, \tilalpha^{a}_{n}$ $(n>0)$ on the right.
Further pushing the Virasoro generators to the oscillator ground states
in $G^{+}_{12}$ or $(G^{+}_{12})^{\dagger}$
using $e^{\alpha_{n}^{a}} q^{L_{0}^{a}} = q^{L_{0}^{a}}
e^{q^{n}\alpha_{n}^{a}}$, and
commuting the creation and annihilation operators between $G^{+}_{12}$ and
 $(G^{+}_{12})^{\dagger}$,
one finds that 
\eqb
 J & = & \prod_{n} \int \frac{d^{2}\vec{z}_{n}}{\pi^{2}} \int \frac{d^{2}\vec{w}_{n}}{\pi^{2}}
       e^{- \vec{z}_{n}\cdot \vec{\zbar}_{n} - \vec{w}_{n}\cdot \vec{\wbar}_{n} } 
       \times 
       e^{q^{n}z_{n2}(c \wbar_{n1}+s \wbar_{n2}) 
       +q^{n}(s \zbar_{n1}+c \zbar_{n2})w_{n1}} \nn \\
      && \times \ \prod_{n} 
       e^{-\frac{1}{n} z_{n1} \alpha_{-n}^{1} +(c\zbar_{n1}-s \zbar_{n2})\tilalpha_{-n}^{1}}
    \cdot  G'
    \cdot  \prod_{n}
        e^{-\frac{1}{n}w_{n2}q^{n}\tilalpha_{n}^{1} +(-s \wbar_{n1} +c \wbar_{n2} ) 
        q^{n}\alpha_{n}^{1} }
        \comma 
\eqe
where $z_{ni}$ are the components of $\vec{z}_{n}$ etc., and 
\eqb
   G' \ = \ 
    g_{+}^{2}\sum_{N,M}
         q^{\ep^{R_{1}}_{k_{2}N, k_{1}M}+\ep^{R_{2}}_{k_{1}N, k_{2}M}-\frac{1}{6}}
       \vert k_{2} N, k_{1}M \rangle \langle k_{2}N, k_{1}M \vert  \comma
\eqe
 for $k_{1}k_{2} \neq 0$ with 
\eqb
   \ep^{R}_{n,m} = \Bigl( \frac{n}{2R}\Bigr)^{2} + \bigl( mR\bigr)^{2}  \period
\eqe
{}For $k_{1}k_{2} = 0$, we have different expressions of $G'$ due to the 
change of the zero-mode structure in $G^{+}_{12}$. We separately discuss this case later.
We then take the $K$-th power of $J$, commute the creation and annihilation operators, and 
perform the $z_{n2}$- and $w_{n1}$-integrals so as to maintain the linearity 
of the oscillators. 
Relabeling $z_{n1}, w_{n2}$ as $z_{n}, w_{n}$, we find that
\eqb
  Z(K) = \Tr_{1}\, J^{K} = g_{+}^{2K}
        \sum_{N,M=-\infty}^{\infty} 
       q^{K (\ep^{R_{1}}_{k_{2}N, k_{1}M}+\ep^{R_{2}}_{k_{1}N, k_{2}M})} 
       q^{-K/6}\prod_{n=1}^{\infty} P_{n} \comma
\eqe
where
\eqb
  P_{n} & = & D_{n}^{K} \prod_{k=1}^{K} \int \frac{ d^{2} z_{n}^{(k)} }{\pi} 
      \int \frac{ d^{2}w_{n}^{(k)} }{\pi} \ e^{- z_{n}^{(k)} \zbar_{n}^{(k)} - w_{n}^{(k)} 
      \wbar_{n}^{(k)}} \nn \\
      && \qquad \times \ e^{s^{2}q^{2n}D_{n}(w_{n}^{(k)}\wbar_{n}^{(k+1)} 
      +\zbar_{n}^{(k)} z_{n}^{(k+1)}) 
      - c q^{n}(1-q^{2n})D_{n}(w_{n}^{(k)}\zbar_{n}^{(k+1)}+\wbar_{n}^{(k)}z_{n}^{(k+1)})} 
      \comma \label{Pn}
\eqe
with $D_{n} = (1-c^{2}q^{2n})^{-1}$ and $ z_{n}^{(K+1)} 
= z_{n}^{(1)}, \, w_{n}^{(K+1)} = w_{n}^{(1)}$.

Since 
\eqb
   \ep^{R_{1}}_{k_{2}N, k_{1}M}+\ep^{R_{2}}_{k_{1}N, k_{2}M} = 
    \Bigl( \frac{k_{2}N}{2 R_{1}\sin \theta_{+} }\Bigr)^{2} 
    + \Bigl( \frac{k_{1}R_{1}M}{\cos \theta_{+}}\Bigr)^{2}
    \comma \label{ee}
\eqe
the sum over $N,M$ gives a product of the theta function  $\vartheta_{3}$. 
The remaining $P_{n}$ are evaluated as follows. 
Introducing a $4K$-vector 
${}^{t}\!\vec{v}= (\Re z_{n}^{(1)}, \Im z_{n}^{(1)}, \Re w_{n}^{(1)}, \Im w_{n}^{(1)}, \cdots)$,
the exponent in $P_{n}$ is expressed as $- {}^{t}\!\vec{v} \cdot M_{K} \cdot \vec{v}$, 
where $M_{K}$ is a $4K \times 4K$ symmetric matrix
\eqb
   M_{K} = \lb 
                 \begin{array}{cccccc}
                    1_{4} & C & & & \cdots & \tC \\
                    \tC  & 1_{4}& C & & \cdots  & \\
                    & \tC & 1_{4} & C & \cdots & \\
                     &&&\ddots && \\
                    &&&& \ddots & \\
                    C & &&&\tC & 1_{4}
                 \end{array} 
              \rb \ \, {\rm with} \ \, 
              C = \lb 
                 \begin{array}{cc}
                    a  (1_{2} -\sigma^{2}) & 0 \\
                    b  \cdot 1_{2} & a  (1_{2} + \sigma^{2})
                 \end{array} 
              \rb  \period
\eqe
$1_{n}$ is the $n \times n$ unit matirx, $\sigma^{2} $ is a Pauli matrix, and 
$a = - s^{2} q^{2n} D_{n}/2$, $b = c q^{n}(1-q^{2n}) D_{n}$. 
Performing the Gaussian integrals then gives  $P_{n} = D_{n}^{K} [ \det M_{K} ]^{-1/2}$. 
The determinant of $M_{K}$ here is regarded as a 
generalization of the circular determinant (see e.g. \cite{GR}), and obtained similarly:
\eqb
    \det M_{K} & = & \prod_{k=1}^{K} \det (1+\omega_{k} C +\omega_{k}^{-1} \tC)
     \nn \\
    &=&  D_{n}^{2K}\prod_{k=1}^{K} \bigl[1-2(c^{2}+d_{k}s^{2})q^{2n} + q^{4n} \bigr]^{2}
     \label{detMK}  \\
    &=& D_{n}^{2K} \bigl[ \, (p_{n}^{+})^{K} - (p_{n}^{-})^{K} \, \bigr]^{4} \comma \nn
\eqe
where $\omega_{k} = e^{2\pi i k/K}$, $d_{k} = \cos(2\pi k/K)$ and
$p_{n}^{\pm} =  (1/2)\bigl[ \sqrt{1-2(c^{2}-s^{2})q^{2n}+ q^{4n}} \pm (1-q^{2n})\bigr]$. 
To derive the last line, we have used a formula (\ref{prodxycos}).
The above expression shows that $\det M_{K}$ and hence $Z(K)$
are actually analytic in $K$.
We also notice that $\prod P_{n}$ gives rise 
to a product of the theta function 
$\vartheta_{1}$. With the help of a formula (\ref{prodsine}), we finally obtain 
\eqb
   Z(K) = g_{+}^{2K} |s|^{K-1} K  
   \vartheta_{3}\Bigl( \frac{itKk_{2}^{2}}{2R_{1}^{2}\sin^{2}\theta_{+}}\Bigr) 
        \vartheta_{3}\Bigl( \frac{2itKk_{1}^{2}R_{1}^{2}}{\cos^{2}\theta_{+}}\Bigr)
     \eta^{K-3}(2it)  \prod_{k=1}^{K-1}
   \vartheta_{1}^{-1}(\nu_{k}|2it) \comma \label{ZK}
\eqe
for $k_{1}k_{2} \neq 0$, where $\eta(\tau)$ is the Dedekind eta function, and 
\eqb
   \pi \nu_{k} = \arcsin\Bigl( |s| \sin \frac{\pi k}{K}\Bigr) \period
   \label{nuk}
\eqe
The $\vartheta_{1}$ part is similar to the oscillator part of the amplitude between D-branes
at angles. This is naturally understood, once we notice that, in the $K$-sheeted Riemann
surface $\calR_{K}$, the array of the interfaces resembles
pairs of D-branes at angles.

\par\medskip\par\medskip
\msubsection{case of $I^{-}_{12}$}

The case of $I^{-}_{12}$ is similar. For the oscillator part, it turns out that  the integral
expression for $P_{n}$ is obtained  by replacing in (\ref{Pn})
$c=\cos 2 \theta_{+}, s = \sin 2 \theta_{+}$ with $-\cos 2 \theta_{-}, \sin 2\theta_{-}$, 
and thus the final expression by  
$\theta_{+} \to \theta_{-}$, i.e., $R_{1} \to 1/2R_{1}$. For the zero-mode part, 
the expression corresponding to (\ref{ee}) is also obtained by $R_{1} \to 1/2R_{1}$.
Therefore, $Z(K)$ in this case is obtained from (\ref{ZK}) by
$R_{1} \to 1/2R_{1}$ (and hence $\theta_{+} \to \theta_{-}$). This is expected, 
since $I^{-}_{12}$ is constructed from the boundary state in which $\CFT_{1}$ is T-dualized 
compared with the boundary state for $I^{+}_{12}$. 
\par\medskip\par\medskip
\msubsection{special cases}

So far, we have considered the case of $k_{1}k_{2} \neq 0$. When $k_{1} k_{2} = 0$, 
while the analysis of the oscillator part remains the same, 
the zero-mode structure and the product of  $G^{\pm}_{12}$ and $(G^{\pm}_{12})^{\dagger}$ change.
Repeating similar computations, one then finds for $I^{+}_{12}$ that 
the product of $\vartheta_{3}$'s in (\ref{ZK}) is replaced by 
$\Theta^{K}_{1} \equiv \vartheta_{3}^{K}\bigl({itk_{1}^{2}}/{2R_{2}^{2}}\bigr) \vartheta_{3}^{K}\bigl(2itk_{1}^{2}R_{1}^{2}\bigr)$ for $k_{1} \neq 0, k_{2}=0$
and $\Theta^{K}_{2} \equiv \theta_{3}^{K}\bigl({itk_{2}^{2}}/{2R_{1}^{2}}\bigr)
 \theta_{3}^{K}\bigl(2itk_{2}^{2}R_{2}^{2}\bigr)$  for 
 $k_{1} = 0, k_{2}\neq 0$. 
When both $k_{1}$ and $k_{2}$ vanish, the original boundary states and hence
the interfaces are not well-defined, since Cardy's condition is not satisfied.
We will not discuss this case. 
From (\ref{kRtheta}), one also finds that $k_{1} k_{2} = 0$ implies $s = 0$ (unless
taking the decompactified limit (or its T-dual) $R_{1,2} = 0,\infty$
which is not covered in our setting). $Z(K)$ is then simplified
as $Z(K) = g_{+}^{2K} \Theta^{K} \eta^{-2K}(2it) $, 
where $\Theta = \Theta_{1}$ or $\Theta_{2}$. 
The results for $I^{-}_{12}$ is obtained by $R_{1} \to 1/2R_{1}$ as above.

\newpage
\msection{Entanglement entropy}

Given the partition functions with the interfaces inserted, we would now like to
discuss the entanglement entropy.
In the following, we concentrate
on the case of $I^{+}_{12}$, since the results for  $I^{-}_{12}$ are easily 
read off from those for $I^{+}_{12}$. We  also focus on  the case 
with $k_{1}k_{2} \neq 0$, unless otherwise stated.

To compute the entropy via (\ref{SZK}), we need the analytic form of  $Z(K)$
in $K$. A way to obtain it is to continue the product in (\ref{ZK}) with respect to $K$, 
and another is to use the last expression in (\ref{detMK}) in terms
of $p_{n}^{\pm}$. 
We first adopt the former with the help  of the Bernoulli polynomials and numbers.
This provides a rather general method to carry out the replica trick.
We then use the latter, which is more straightforward. In the course of 
showing the equivalence of the results from the two approaches, we find that the entropy 
is expressed by the dilogarithm function.

We start  with the result of $Z(K)$ in (\ref{ZK}). 
Since the modular parameter $t = \pi/(2\log L)$ is small for $L \gg 1$, 
it is convenient to evaluate it by
the modular transformation $\tau \to -1/\tau$.
One then finds that (when $|s| \neq 0$)
\eqb
   Z(K) 
    & = & \frac{(g_{+}^{2} |s|)^{K}}{|k_{1}k_{2}|} \,  e^{-(K-3)\pi/24 t} \, 
       e^{\varphi(K)/t} \Bigl( 1 + \calO(e^{-\mu /t}) \Bigr) \comma \label{ZKasympt}
\eqe
where $\mu$ is a positive constant and
\eqb
   \varphi(K)  =  \frac{\pi}{2} \sum_{k=1}^{K-1} \Bigl( \half - \nu_{k}\Bigr)^{2}
     \comma
\eqe
with $0 < \nu_{k} < 1$ $(k= 1, ..., K-1)$. Note that $\varphi(K)$ is of the form
\eqb
   \varphi(K)=\sum_{k=1}^{K-1}f\Bigl(\frac{k}{K}\Bigr) \comma
   \quad f(x)=\frac{1}{2\pi}\arccos^2(|s|\sin\pi x)\period
\eqe
Since $f(x)$ is analytic around $x=0$, we expand it as  $f(x)=\sum_{m=0}^\infty f_mx^m$.
A useful fact here is that 
 $\sum_{k} k^{m}$ is expressed by the Bernoulli polynomials $b_{n}(x)$
and numbers $b_{n}$ as in (\ref{km}). 
{}From this and properties of $b_{n}(x), b_{n}$
summarized in Appendix, it follows that
\eqb
   \partial_K\varphi(K)\Big|_{K=1}
& = & \sum_{m=0}^\infty \frac{f_m}{m+1} \del_{K} b_{m+1}(K) \Big|_{K=1} \nn \\
&=& f(0)+\frac{1}{2}f'(0)
 +\int_0^\infty \frac{if'(ix)- if'(-ix)}{1-e^{2\pi x}} dx
\period
\eqe 
After plugging the explicit form of $f(x)$ and changing the variables as
$u=$
\linebreak[5]
$\mbox{arcsinh}\,(|s|\sinh\pi x)$, we apply the result to (\ref{SZK}), and obtain
\eqb
   S = \sigma(|s|)  \log L 
        - \log|k_{1}k_{2}| \comma    \label{S}
\eqe
up to terms vanishing for $L \gg 1$,  where
\eqb
\sigma(|s|) = \frac{|s|}{2} - \frac{2}{\pi^{2}}  
  \int_0^\infty  u
\Bigl(\sqrt{1+ \bigl(|s|/\sinh u\bigr)^{2}}-1\Bigr)du
   \period \label{sigmas}
\eqe
We find that the entropy has a logarithmic scaling with respect to the size of
the system $ L $, but the coefficient $\sigma(|s|)$ is a function of $|s|$.
It turns out shortly that $\sigma(|s|)$ is expressed  by the dilogarithm function.
The sub-leading term counts the product of the winding numbers.
This is analogous to the topological entanglement entropy in (2+1)-dimensional
systems characterizing the topological order 
\cite{Kitaev:2005dm,LW}.
We also note that the entropy is a function of $\theta_{+}, k_{1}, k_{2}$ only, and
does not depend on $\alpha,\beta, R_{1}, R_{2}$ explicitly.

In special cases, $\sigma(|s|)$ is simplified. 
First, let us consider the topological case, $|s| =1$.
Since the identity is included as a special case, this case should 
reproduce the universal 
scaling of the ordinary entanglement entropy without interfaces. In fact, 
one finds that $\sigma(1) = c/3 = 1/3$, which agrees with the result 
\cite{Holzhey:1994we,Calabrese:2004eu}.
Next, when $|s|$ is small,
one can show that the second term in (\ref{sigmas}) is $\calO(|s|^{2}\log |s| )$, and that
$\sigma(|s|) \to |s|/2$. This implies that the leading term decreases  as $|s|$
does,  which also agrees with the fact that the oscillator part of the two CFT's are decoupling 
as $|s| \to 0$.
The result for small $|s|$ is derived also by directly expanding $\nu_{k}$ in (\ref{nuk}).
For general $|s|$, one can check that $\sigma(|s|)$ monotonically interpolates
these two cases.  This supports an intuition that
the entanglement changes according to $|s|$, 
since $|s|$ is the strength of the interaction between the two CFT's.
In \cite{Quella:2006de},  certain reflection and  transmission
coefficients are  introduced as probes
of conformal interfaces. For the $c=1$ permeable interfaces, they give $c^{2}$ and $s^{2}$.
Compared with those coefficients, one finds that 
the entanglement entropy (\ref{S}) can probe
a little more details of the interfaces. 

When the sizes of the two systems are $L$ and $\Lambda -L$,
the entanglement entropy without interfaces scales as
$(c/3)\log\bigl[(\Lambda/\pi) \sin (\pi L/\Lambda)\bigr] \, + \, $const. \cite{Holzhey:1994we},
where the sub-leading constant term is independent of $L$
\cite{Calabrese:2004eu}.
In our case with interfaces, 
the entropy should also be symmetric under the exchange of the two CFT's, and 
the above scaling should be reproduced in a special case.
A  possible form for  $L \neq\Lambda/2 $ satisfying
these requirements is $S = \sigma(|s|)\log\bigl[(\Lambda/2) \sin (\pi L/\Lambda)\bigr]
-\log|k_{1}k_{2}|$.

Here, some comments for special cases may be in order. 
When $k_{1} k_{2} = 0$ and hence $s =0$, it follows from the result in section 3.3
that the entropy exactly vanishes: $S = 0$. This confirms the fact that the two CFT's decouple
in this case.
When $k_{1}k_{2} \neq 0$ and hence
$s \neq 0$ (unless in the decompactified limit), 
the entropy might appear to be negative for small enough $s$. This, however, is not the case: 
To obtain (\ref{ZKasympt}) by modular transformations, we have used $t/s^{2} \ll 1$ for $\vartheta_{3}$'s.
Thus, the result in (\ref{S}) is valid when the first term is large enough.
In fact, since $L$ is the cutoff in our setting 
and can be arbitrarily large independently of other parameters,  
this condition is always satisfied  by taking large enough $L$. 
Note that, however small $s$ is, the two CFT's couple through the zero-modes 
if $k_{1}k_{2} \neq  0$.
In order to analyze the case where $k_{1} k_{2} \neq 0$ 
and $s \log L $ is small with actually finite $L$,
one may need to develop a method to compute 
 the entanglement entropy for finite systems, e.g., by generalizing the results 
in \cite{Calabrese:2004eu}. 

One can also  derive the entropy  by applying the expression of 
$Z(K) =$
\linebreak[5]
$g_{+}^{2K} \vartheta_{3}\vartheta_{3} q^{-K/6} \prod P_{n}$
in terms of $p_{n}^{\pm}$.
Recalling the formula (\ref{SZK}), we first evaluate
\eqb
   \sum_{n=1}^\infty\partial_K\log P_n \, \Big\vert_{K=1} &=&
   -2 \sum_{n=1}^\infty \, F(2\pi t n)
   \nn \\
   & \simeq & \frac{-1}{\pi t} \int_{0}^{\infty} F(y) \, dy + F(0) 
   \comma
\eqe
as $t \to 0$, where
\eqb
  F(y) = \log|s|-y+\sqrt{1+\bigl(|s|/\sinh y\bigr)^2}\, 
\mbox{arcsinh}\,\Bigl(\sinh y/|s|\Bigr)  \comma
\eqe
and $F(0) = 1+ \log |s| $.
We have used $p_{n}^{+}p_{n}^{-} = |s|^{2} q^{2n}$, and assumed that
$s$ is not vanishing so that $F(0)$ is not divergent.
In the case without the derivative $\del_{K}$, 
a similar approximation by an integral is subtle, since  the summand 
is singular at $t=0$. Thus, we instead note that  
$\prod_{n=1}^{\infty} P_{n} \, \big\vert_{K=1} = e^{-\pi t/3} \eta^{-2}(2it)$, 
which after a modular transformation gives  
$ \sum_{n=1}^{\infty} \log P_n \, \big\vert_{K=1} \simeq \pi/(12 t) + \log(2t)$.
Together with  $\vartheta_{3} \vartheta_{3} \simeq |s|/(2Kt|k_{1}k_{2}|) $
for small $t$, the entropy is obtained 
as $S = \tilde{\sigma}(|s|) \log L - \log|k_{1}k_{2}|$, with
\eqb
 \tilde{\sigma}(|s|) = \frac{1}{6} + \frac{2}{\pi^{2}} \int_{0}^{\infty} F(y) \, dy 
 \period
\eqe

Compared with the previous result (\ref{S}), $\tilde{\sigma}(s)$ should agree with 
$\sigma(s)$. To show this, we consider their derivatives:
\eqb
  \sigma'(s)= \frac{1}{2}-\frac{2}{\pi^2}\int_0^\infty  \!
\frac{dw \ \mbox{arcsinh}\,w}{w\sqrt{1+w^2}\sqrt{1+w^2/s^2}} \comma
 \ \tilde\sigma'(s)= \frac{2}{\pi^2}\int_0^\infty \!
\frac{dz \ \mbox{arcsinh}\,z}{z\sqrt{1+z^2}\sqrt{1+s^2z^2}} \comma
\eqe
and
\eqb
 \sigma''(s)=\frac{-2}{\pi^{2}s}\int_0^\infty \!
\frac{dw\ (w/s^2)\, \mbox{arcsinh}\,w}{\sqrt{(1+w^2)(1+w^2/s^2)^3}} \comma
\  \tilde\sigma''(s)=\frac{-2}{\pi^{2}s}\int_0^\infty \!
\frac{dz \ (s^2 z) \, \mbox{arcsinh}\,z}{\sqrt{(1+z^2)(1+s^2 z^2)^3}} \comma \nn \\
\label{d2sigma0}
\eqe
where we have made changes of variables $w=\sinh u$ and $z= s^{-1}\sinh y$.
The integral for $\sigma''(s)$ here is performed as
\eqb
   -\frac{\pi^{2}s}{2} \sigma(s)''& =& \frac{1}{s-1/s}
\Biggl(\sqrt{\frac{w^2+1}{w^2+s^2}}\, \mbox{arcsinh}\,w
-\mbox{arcsinh}\frac{w}{s}\Biggr)\Bigg|_{w=0}^\infty\nn\\
&=&\frac{\log s}{s-1/s} \period \label{d2sigma}
\eqe
One then finds that $-(\pi^{2}s/2) \tilde{\sigma}(s)''$ is also given by 
the above, 
namely $ \sigma(s)'' = \tilde{\sigma}(s)'' $, 
since the integral representations of $ s \cdot \sigma''$ and 
$s \cdot \tilde{\sigma}''$  are related by 
$ s \leftrightarrow 1/s$.
It is easy to confirm that 
the integration constants are  also the same, e.g., by checking special values 
$\sigma(1) = \tilde{\sigma}(1) = 1/3, \, \sigma'(1) = \tilde{\sigma}'(1) = 1/4$, 
which verifies $\sigma(s) = \tilde{\sigma}(s)$.

As a by-product, we find by integrating $\sigma''(s)$ that $\sigma(s)$
is expressed as
\eqb
  \sigma(s)=\frac{1}{6}+\frac{s}{3}+\frac{1}{\pi^2}
\Bigl[ (s+1)\log(s+1)\log s
+(s-1)\, \mbox{Li}_2(1-s)+(s+1)\, \mbox{Li}_2(-s)\Bigr]
\comma
\eqe
where $\mbox{Li}_2(z)$ is the dilogarithm function. 
We summarize some properties of $\mbox{Li}_2(z)$ in Appendix.
Using them, one can rederive the values of $\sigma(1)$,
$\sigma'(1)$, and the small-$s$ behavior of $\sigma(s)$.

\msection{Summary}
We have obtained  the partition functions with the  $c=1$ permeable
interfaces inserted, and the entanglement entropy of the corresponding
interface CFT analytically. 
The entropy scales logarithmically with respect 
to the size of the system, as in the case without interfaces 
\cite{Holzhey:1994we,Calabrese:2004eu}. Its coefficient, however,
is not a constant but a monotonic function of $|s|$ controlling the permeability,
and is given explicitly in terms of the dilogarithm function.
The sub-leading term of the entropy 
counts the product of the winding numbers.
This is analogous to the topological entanglement entropy, which
characterizes the topological order in $(2+1)$-dimensional systems \cite{Kitaev:2005dm,LW}.

Our results show that the entanglement entropy is a useful probe 
to the system, as in the case without interfaces. It would be interesting to 
study how general our findings are: For example, does the entropy always
contain the topological  information of the system? Does it always 
show the scaling as in the case without interfaces? 
Regarding such studies, it would be useful to generalize
the analysis based on the conformal symmetry 
\cite{Holzhey:1994we,Calabrese:2004eu,Cardy:1988tk} 
to the case with interfaces. A complication with interfaces is that one has to 
keep track of the shape of  interfaces under conformal transformations. 
It would also be interesting to consider implications
of the entanglement entropy in the context of condensed matter
physics and string theory.

%
\vspace{3ex}
\begin{center}
  {\bf Acknowledgments}
\end{center}

We would like to thank YITP
where  part of this work was carried out during the workshop 
``Development of quantum field theory and string theory'', 
and T. Takayanagi for useful correspondences.
 Y.S. would also like to thank the participants of the workshop 
 ``Liouville, integrability and branes (4)'' at APCTP
 for bringing his attention to conformal interfaces, and 
 Y. Hikida and N. Ishibashi for useful conversations and discussions. 
Research of K.S. is supported in part by
Grant-in-Aid for Scientific Research
from the Japan Ministry of Education, Culture, Sports, Science
and Technology and by Keio Gijuku Academic Development Funds.

\appsection{Appendix}

In the main text, we apply the formulas 
\eqb
  && \prod_{r=0}^{n-1}
\Bigl[x^2-2xy\cos\bigl(\theta+\frac{2r\pi}{n}\bigr)+y^2\Bigr]
=x^{2n}-2x^n y^n\cos n\theta +y^{2n} \comma \label{prodxycos} \\ 
&& \prod_{r=1}^{n-1} \sin \Bigl(\frac{r \pi}{n}\Bigr) =  \frac{n}{2^{n-1}} \period
  \label{prodsine} 
\eqe
We also use the Bernoulli polynomials (see e.g. \cite{GR}) $b_{n}(x)$ ($n = 0,1,2, ...$) 
defined by 
\eqb
   \frac{t e^{xt}}{e^{t}-1} = \sum_{n=0}^{\infty} b_{n}(x) \frac{t^{n}}{n!} \quad \
   (|t| < 2\pi)
   \period
\eqe
Their derivatives are  $ b'_{n}(x) = n b_{n-1}(x) $.
At $x=0$, they give the Bernoulli numbers $b_{n}$, namely,  $b_{n} = b_{n}(0)$.
$b_{n}$ with odd index vanish except for $b_{1}$, and 
$b_{0} = 1, b_{1} = -1/2, b_{2} = 1/6, b_{4} = -1/30, ...$ . 
One also has $b_{n} = b_{n}(1)$ for $n \neq 1$, and $b_{1}(1) = 1/2$.
By the Bernoulli polynomials, the sums of
powers of natural numbers are expressed as
\eqb
   (m+1) \sum_{k=1}^{n-1} k^{m} = b_{m+1}(n) - b_{m+1} \quad \ (n,m = 1,2, ...)
   \period \label{km}
\eqe
The Bernoulli numbers with even index have an integral representation,
\eqb
  b_{2n}=  4n (-1)^{n} \int_0^\infty\frac{t^{2n-1}}{1-e^{2\pi t}} dt \period
  \label{bs}
\eqe 
{}From this, it follows that 
\eqb
 \frac{1}{m+1} \del_{n} b_{m+1}(n) \Big|_{n=1} = 
 \delta_{m,0}+\frac{1}{2}\delta_{m,1}
+\left(i^m-(-i)^m\right)\int_0^\infty\frac{mt^{m-1}}{1-e^{2\pi t}}dt
\period
\eqe

In section 4, we use the dilogarithm function defined by
\eqb
\mbox{Li}_2(z)=\sum_{k=1}^\infty\frac{z^k}{k^2}
=-\int_0^z\frac{\log(1-w)}{w}dw \period \label{Li2}
\eqe
$\sigma''(s)$ in (\ref{d2sigma}) is integrated by using the above integral representation
and
\eqb
  \int dz \, \mbox{Li}_2(z) = z \, \mbox{Li}_2(z) - z - (1-z) \log(1-z) \period
\eqe 
{}From $\mbox{Li}_2(1) = \pi^{2}/6$, $\mbox{Li}_2(-1) = -\pi^{2}/12$, 
one can check the values of $\sigma(1)$ and $\sigma'(1)$. To derive
the small-$s$ behavior of $\sigma(s)$, useful formulas are (\ref{Li2})
and
\eqb
   \mbox{Li}_2(1-z) = -\mbox{Li}_2(z) - \log z \log(1-z) + \frac{\pi^{2}}{6}
   \period
\eqe

%
%
\def\thebibliography#1{\list
 {[\arabic{enumi}]}{\settowidth\labelwidth{[#1]}\leftmargin\labelwidth
  \advance\leftmargin\labelsep
  \usecounter{enumi}}
  \def\newblock{\hskip .11em plus .33em minus .07em}
  \sloppy\clubpenalty4000\widowpenalty4000
  \sfcode`\.=1000\relax}
 \let\endthebibliography=\endlist
%
%
\vspace{3ex}
\begin{center}
 {\bf References}
\end{center}
\par \vspace*{-2ex}

%

%
\end{document}